\documentclass[conference]{IEEEtran}
\IEEEoverridecommandlockouts
\usepackage[table]{xcolor}
\usepackage{cite}
\usepackage{graphicx}
\usepackage{lipsum}
\usepackage{stfloats}
\usepackage{amssymb}
\usepackage{pifont}
\usepackage{bbding}
\usepackage{booktabs}
\usepackage{ctable}
\usepackage{threeparttable}
\usepackage{footmisc}
\usepackage{framed} 
\usepackage{array}
\usepackage{multirow}
\usepackage{longtable}
\usepackage{rotating}
\usepackage{fancyhdr}
\usepackage{lastpage}
\usepackage{layout}
\usepackage{wrapfig}
\usepackage{xcolor}
\usepackage{tabularx}
\usepackage{tabu}
\usepackage{enumitem}
\usepackage{url}
\usepackage{amsmath}
\graphicspath{{Figures/}}
\DeclareGraphicsExtensions{.png, .eps}

\begin{document}
	\title{Wireless Resource Management in Intelligent Semantic Communication Networks}
	\author{\IEEEauthorblockN{Le Xia\IEEEauthorrefmark{1},
	Yao Sun\IEEEauthorrefmark{1}\thanks{Corresponding author: Yao Sun (Yao.Sun@glasgow.ac.uk).},
	Xiaoqian Li\IEEEauthorrefmark{2},
	Gang Feng\IEEEauthorrefmark{2}, and
	Muhammad Ali Imran\IEEEauthorrefmark{1}}
	\IEEEauthorblockA{\IEEEauthorrefmark{1}James Watt School of Engineering,
	University of Glasgow, Glasgow, UK\\
	\IEEEauthorrefmark{2}National Key Lab on Communications, University of Electronic Science and Technology of China, Chengdu, China\\
	Email: l.xia.2@research.gla.ac.uk, \{Yao.Sun, Muhammad.Imran\}@glasgow.ac.uk, \{xqli, fenggang\}@uestc.edu.cn}}
	
	
	\maketitle
	\begin{abstract}
	The prosperity of artificial intelligence (AI) has laid a promising paradigm of communication system, i.e.,~\textit{intelligent semantic communication} (ISC), where semantic contents, instead of traditional bit sequences, are coded by AI models for efficient communication.
	Due to the unique demand of background knowledge for semantic recovery, wireless resource management faces new challenges in ISC. 
	In this paper, we address the user association (UA) and bandwidth allocation (BA) problems in an ISC-enabled heterogeneous network (ISC-HetNet).
	We first introduce the auxiliary knowledge base (KB) into the system model, and develop a new performance metric for the ISC-HetNet, named system throughput in message (STM).
	Joint optimization of UA and BA is then formulated with the aim of STM maximization subject to KB matching and wireless bandwidth constraints. 
	To this end, we propose a two-stage solution, including a stochastic programming method in the first stage to obtain a deterministic objective with semantic confidence, and a heuristic algorithm in the second stage to reach the optimality of UA and BA.
	Numerical results show great superiority and reliability of our proposed solution on the STM performance when compared with two baseline algorithms.
	\end{abstract}
	

	\IEEEpeerreviewmaketitle
	
	\section{Introduction}
	Artificial intelligence (AI) technology has been widely regarded as an indispensable component in future networking paradigms.
	Benefited from a variety of state-of-the-art deep learning (DL) techniques, many sophisticated computation tasks can be well accomplished.
	Moreover, due to the limited wireless resources, traditional communication system is becoming gradually insufficient to process diversified service requirements under various application scenarios.
	This destined bottleneck is, therefore, motivating us to hunt for a bold change in new designs on AI-enabled 6G networks, for a paradigm revolution from traditional bit-based communication to~\textit{intelligent semantic communication} (ISC)~\cite{guler2018semantic, xie2021deep, xie2020lite, weng2021semantic}.
	
	The concept of semantic communication was first defined in Shannon and Weaver's landmark paper~\cite{shannon1948mathematical}.
    Different from traditional communication system, semantic communication focuses on distilling core semantics by coding source messages (i.e., semantic encoding), to filter out those contents irrelevant to the predefined conveyed information, thus greatly reducing the amount of transmitted data while reserving original semantics.
    Meanwhile, powerful interpreters deployed in the destination devices can recover true meanings from received signals (i.e., semantic decoding), even if intolerable bit errors may exist at syntactic level~\cite{strinati20216g}.
    Accordingly, semantic communication can significantly save the required bandwidth and guarantee adequate communication reliability, which thus becomes a preeminent actor to drive 6G networks a dramatic leap forward.
    In parallel, emerging AI technology under fast development has revealed its great potential in diverse future-oriented wireless communication scenarios~\cite{xia2021smart}, which also renders a viable path to undertake such semantic coding tasks for ISC.
    For instance, as studied in~\cite{xie2021deep} and~\cite{xie2020lite}, text-based ISC becomes promising by leveraging a natural language processing (NLP) model called Transformer.
    Besides, powered by the attention mechanism, a squeeze-and-excitation network was employed in~\cite{weng2021semantic} to achieve speech-based ISC.
    
    Although there are several preliminary research investigations on ISC from a link-level perspective, challenges on the upper network layer have not been explored yet, including user association (UA) and bandwidth allocation (BA) problems among multi-tier base stations (BSs) in an ISC-enabld heterogeneous network (ISC-HetNet).   
	Notably, conventional solutions to UA and BA cannot be directly extended to the ISC-HetNet scenario, due to some inevitable changes on both network architecture and communication system along with additional limitations.
	Specifically, a proper semantic-related metric should be defined to assess the system performance of ISC-HetNet.
	In the meantime, multiple mobile users (MUs) should be correctly associated with specific BSs to match their required background knowledge, which also brings new semantic constraints.
	Hence, the exploration for new UA and BA strategies becomes particularly meaningful in the context of ISC-HetNets, and to the best of our knowledge, no research article has addressed this problem before.
	
	In this paper, we study the UA and BA problems in the downlink of a multi-tier ISC-HetNet.
    According to the specific characteristics of ISC, a two-stage solution is proposed to reach the optimality of semantic performance.
    In a nutshell, this paper has the following contributions.
    \begin{itemize}
		\item \textit{Semantic model construction:} We first introduce auxiliary knowledge base systems for BSs, which is to ensure necessary background knowledge for semantic reconstruction of the associated MUs. Besides, the relationship between message-rate and bit-rate is elucidated, thereby developing a new performance metric for the ISC-HetNet, named system throughput in message (STM).
		\item \textit{Network problem formulation:} Due to the dynamic background knowledge matching degrees between different MUs and BSs, we define knowledge matching coefficients for ISC-HetNet to formulate a stochastic optimization problem to jointly solve UA and BA with the aim of maximizing STM.
		\item \textit{Two-stage solution:} We propose a two-stage solution to the devised problem. In the first stage, the primal stochastic problem is converted into a deterministic one by the semantic confidence level introduced, followed by the second stage in which we exploit an interior-point method and a heuristic algorithm to find the optimal UA and BA solution, respectively. 
		\item \textit{Simulation verification:} Simulations are conducted in a text-based ISC-HetNet, where a Transformer model is utilized to complement the semantic coding part. Numerical results demonstrate that our proposed solution can smoothly complete UA and BA tasks, and outperform two max-SINR-based algorithms on the STM performance.		
	\end{itemize}
    
    The remainder of this paper is organized as follows.
    The next section presents the whole ISC-HetNet model along with the newly defined STM performance metric. 
    In Section III, the semantic-based stochastic problem is formulated with extra semantic constraints.
    Then the two-stage solution is described in Section IV.
    In Section V, we finally present our numerical results before concluding this paper in Section VI.    
	
	\section{System Model}
	The focus of this paper is on the downlink of the ISC-HetNet, where all BSs can provide their associated MUs with ISC services.
	In parallel, as elucidated in~\cite{xie2021deep},~\cite{strinati20216g}, and~\cite{bao2011towards}, one crucial factor strongly related to the accuracy of semantic interpretation is the knowledge matching degree between source and destination.
	Herein, we define the auxiliary system of~\textit{knowledge base} (KB) that can provide MUs with a specific application domain of background knowledge for their semantic reconstruction.
	Therefore, if the KB at the receiver closely matches the one at the transmitter, ultra-low interpretation errors can be obtained in ISC with sufficient reliability.
	
	\subsection{Network Topology of the ISC-HetNet}
	Suppose that an ISC-HetNet is composed of multiple tiers of BSs, including macro BSs (MBSs), pico BSs (PBSs), and femto BSs (FBSs), where the BSs at the same tier have a uniform and fixed transmit power.
	Here we assume there are $L$ BSs in a set denoted by $\mathcal{B}=\left\{\mathcal{BS}_{1},\mathcal{BS}_{2},\ldots,\mathcal{BS}_{L}\right\}$, where $\mathcal{BS}_{j}$ indicates the $j_{th}$ BS.
	Besides, $M$ MUs are randomly located within the coverage of these BSs.
	Thus, let $\mathcal{U}=\left\{\mathcal{MU}_{1},\mathcal{MU}_{2},\ldots,\mathcal{MU}_{M}\right\}$ denote the set of all MUs, where $\mathcal{MU}_{i}$ represents the $i_{th}$ MU.
	
	Then, let $x_{ij}\in \left\{ 0,1\right\}$ denote the association indicator, where $x_{ij}=1$ means that $\mathcal{MU}_{i}$ is associated with $\mathcal{BS}_{j}$, and $x_{ij}=0$ otherwise.
	Suppose that only single-BS association is allowed at a time in the ISC-HetNet, so we have
	\begin{equation}
		\sum_{j\in \mathcal{B}} x_{ij}= 1,\  \forall i\in \mathcal{U}.\label{singleBS}
	\end{equation}
	
	Apart from the UA, each BS needs to allocate bandwidth resources to its served MUs.
	Here, we define that there is $n_{ij}$ amount of bandwidth assigned to $\mathcal{MU}_{i}$ by $\mathcal{BS}_{j}$, and the bandwidth budget of $\mathcal{BS}_{j}$ is $N_{j}$.
	Hence, $n_{ij}$ should satisfy
	\begin{equation}
		\sum_{i\in \mathcal{U}}x_{ij} n_{ij}\leqslant N_{j},\  \forall j\in \mathcal{B}.
	\end{equation}
	
	Next, let $\gamma_{ij}$ denote the signal-to-interference-plus-noise ratio (SINR) experienced by $\mathcal{MU}_{i}$ from $\mathcal{BS}_{j}$, which is typically averaged and considered a constant when associated.
	According to Shannon's classical information theory, the downlink bit-rate of $\mathcal{MU}_{i}$ served by $\mathcal{BS}_{j}$ with $n_{ij}$ bandwidth is
	\begin{equation}
		b_{ij}=n_{ij}\log_{2}\left( 1+\gamma_{ij}\right).\label{longtermrate}
	\end{equation}
	So the whole system throughput in bit can be phrased as
	\begin{equation}
		T_{B}=\sum_{i\in \mathcal{U}}\sum_{j\in \mathcal{B}}x_{ij}b_{ij}=\sum_{i\in \mathcal{U}}\sum_{j\in \mathcal{B}}x_{ij}n_{ij}\log_{2}\left( 1+\gamma_{ij}\right).\label{bitthroughput}
	\end{equation}
	
	\subsection{KB-enabled Semantic Model}
	
	\begin{figure}[t]
		\centering
		\includegraphics[width=0.49\textwidth]{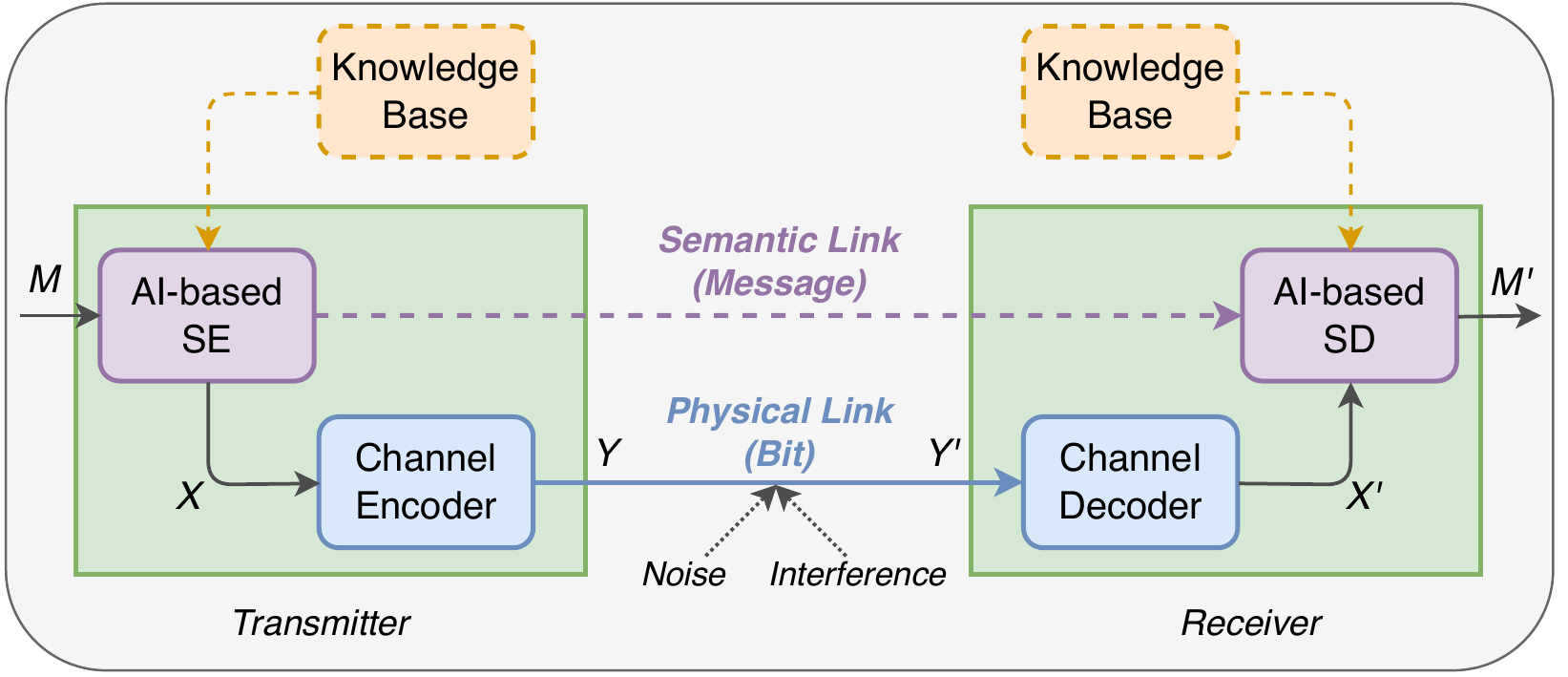} 
		\caption{The structure diagram of an ISC system.}
		\label{SC-System}
    \end{figure}
	
	Fig.~\ref{SC-System} illustrates an ISC system, which introduces an AI-based semantic encoder (SE) and an AI-based semantic decoder (SD) to handle semantic coding tasks.
	To be explicit, the SE is for extracting intrinsic semantic information from the source messages, while the SD is to reconstruct original messages from the received signals.
	Especially from the perspective of a semantic link, the focus of communication becomes the message itself implied in the transmitted bits, with the assistance of both SE and SD.
	It is also worth mentioning that different semantic models may produce distinct amount of bits for the same semantic representation.
	Taking NLP models as exemplification, in the text transmission scenario, a Transformer model is believed to require fewer bits for encoding a given sentence than using a typical word2vec model~\cite{mikolov2013efficient}, as a role of SE.
	Additionally, message property (MP) should be the second primary factor affecting the bit-to-message transformation, including the message type and message length, etc.
	For instance, an image message generally demands more bits for semantic representation when compared with a text message containing the same semantics, and a text message with longer sentences may consume more bits for encoding.
	Moreover, the knowledge matching is the third factor of ISC, i.e., the aforementioned KB system.
	Herein, let $\mathcal{B}_{i}$ denote the set of BSs holding the KB clusters with the most knowledge required for $\mathcal{MU}_{i}$, where $\mathcal{B}_{i}\subseteq \mathcal{B},\ \forall i\in \mathcal{U}$.
	This means the feasible associated BSs of $\mathcal{MU}_{i}$ must be constrained within $\mathcal{B}_{i}$.
	Having this semantic-based KB constraint,~(\ref{singleBS}) can be now updated by
	\begin{equation}
		\sum_{j\in \mathcal{B}_{i}} x_{ij}= 1,\  \forall i\in \mathcal{U}.
	\end{equation}
	
	Since the message itself has become the focus of ISC, a new metric, named as~\textit{system throughput in message} (STM), is apparently more reasonable and applicable when compared to the traditional metric of bit throughput.
	The STM aims to measure the sum of the message-rates that are delivered to all MUs within a time unit.
    Herein, the unit message, in a communicating sense, indicates a complete piece of information successfully transmitted into the semantic communication link.
	As exemplification, an entire text sentence ending with a period in text communication, or a voice signal completely sent out in speech communication, can be regarded as a message.
	With that in mind, the aforementioned~\textit{message-rate} is thus interpreted as the amount of messages that are conveyed or processed per time unit (symbol: $msg/s$), with the reference of the bit-rate definition (symbol: $bit/s$).
	In the following, we define an important transformation relationship between message-rate and bit-rate, since the latter already has a very sound systematic framework based on Shannon's theory.
	
	\begin{figure}[t]
		\centering
		\includegraphics[width=0.25\textwidth]{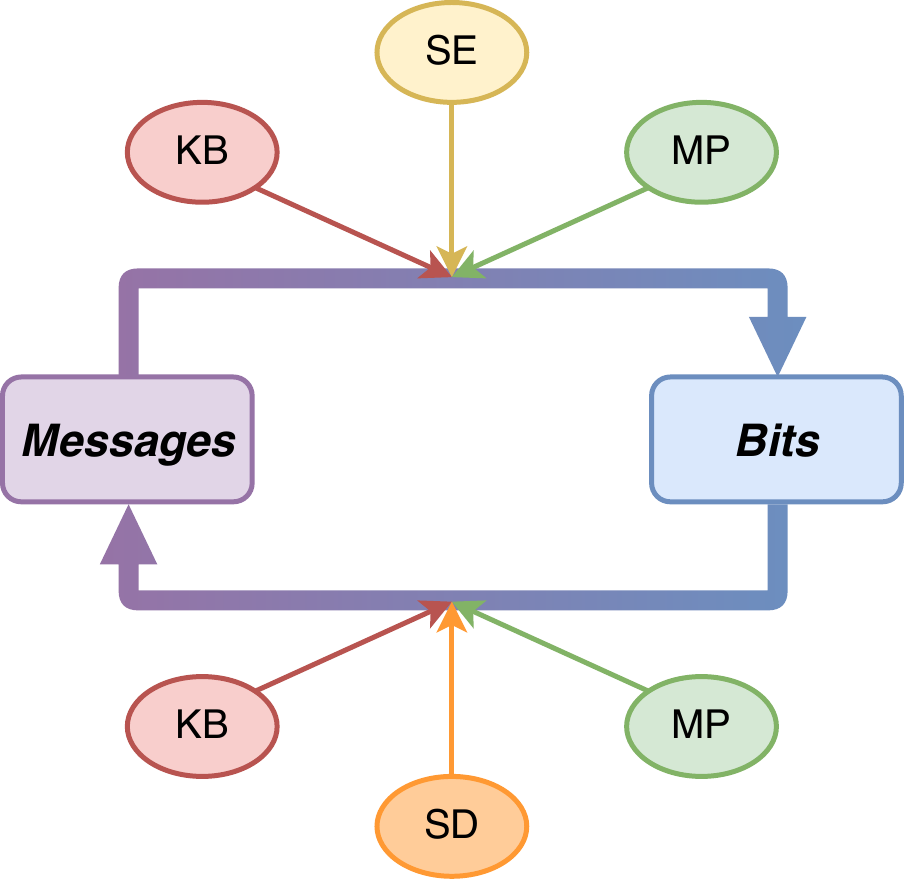} 
		\caption{The transformation diagram between received bits and recovered messages in semantic communication.}
		\label{Conversion}
    \end{figure}
	
	For all $i \in \mathcal{U}$ in the ISC-HetNet, let $S_{i}\left(\cdot \right)$ denote a universal bit-to-message (B2M) transformation function of $\mathcal{MU}_{i}$ under given channel conditions, which is to convert bit-rate to message-rate in a physical sense.
	Note that the manifestation of $S_{i}\left(\cdot \right)$ should be solely from users' side, jointly determined by adopted semantic models, associated KBs, and received MPs, which have been summarized in Fig.~\ref{Conversion}.
	Intuitively, the more bits the transmitter sends, the more messages the receiver gets.
	Hence, $S_{i}\left(\cdot \right)$ related to each user can be always deemed as monotonically increasing and continuously differentiable.
	
	Then in view of the bit-rate $b_{ij}$ given in~(\ref{longtermrate}), we can thus take advantage of $S_{i}\left(\cdot \right)$ to naturally define the message-rate by $\xi_{ij}$, where we put
	\begin{equation}
		\xi_{ij}=S_{i}\left(b_{ij}\right),\ \forall \left( i,j\right) \in \mathcal{U}\times \mathcal{B},
	\end{equation}
	such that
	\begin{equation}
		T_{M}=\sum_{i\in \mathcal{U}}\sum_{j\in \mathcal{B}}x_{ij}\xi_{ij}=\sum_{i\in \mathcal{U}}\sum_{j\in \mathcal{B}}x_{ij}S_{i}\left(b_{ij}\right).
	\end{equation}
	Here $T_{M}$ is the STM of the ISC-HetNet that can well represent the network performance from a semantic perspective.
	
	\section{Problem Formulation}
	In the ISC-HetNet, even if there exists a set of BSs, i.e., $\mathcal{B}_{i}$, holding multiple auxiliary KBs demanded by $\mathcal{MU}_{i}$, a certain degree of~\textit{semantic ambiguity} may still occur in the process of semantic interpretation due to the imperfect matching of background knowledge.
	Specially, the semantics received by MUs may contain knowledge relevant to multiple application domains at the same time, whereas each KB can only offer the knowledge in one domain.
	Therefore, let $S_{i}^{P}\left(\cdot \right)$ denote the B2M function in a case of perfect knowledge matching, in which all required knowledge of $\mathcal{MU}_{i}$ can be provided.
	Herein, $S_{i}^{P}\left(\cdot \right)$ can be reckoned as a strictly convex, monotonically increasing, and continuously differentiable function.
	For instance, in Transformer-based text communication, the transformation ratio between bits and messages is averaged over the association time, so it becomes a constant related to the number of neurons in the input layer of SD, regardless of dynamics in transmitted messages.
	Further, let $S_{i}^{M}\left(\cdot \right)$ indicate a case of maximum knowledge matching, which means that $\mathcal{MU}_{i}$ can only get partial but maximum knowledge in the ISC-HetNet for semantic reconstruction.
	Note that only the case to $S_{i}^{M}\left(\cdot \right)$ will be discussed in this paper as it is a more general scenario in semantic communication\footnote{$S_{i}^{P}\left(\cdot \right)$ is only a special case of the framework we define in this paper.}.
		
	We first assume the amounts and types of different KBs are randomly distributed among all BSs, thereby, possible knowledge gaps exist in the ISC-HetNet.
	Hence, it is undeniable that the number of correctly reconstructed messages could be reduced when compared with that in perfect knowledge matching cases.
	According to this rationale, we define
		\begin{equation}
			S_{i}^{M}\left(\cdot \right)=\eta_{i}\cdot S_{i}^{P}\left(\cdot \right),\ \forall i\in \mathcal{U}.\label{Def4}
		\end{equation}
	Here, $\eta_{i}$ ($0<\eta_{i}<1$) represents the knowledge matching coefficient, and the larger the value of $\eta_{i}$, the higher the knowledge matching degree of $\mathcal{MU}_{i}$.
		
	Notably, the knowledge gap is foreseeable to be highly dynamic due to different user situations, which leads to that $\eta_{i}$ should be a random variable.
	Without loss of generality, in this paper, we assume that $\eta_{i}$s are independent and identically distributed, while satisfying a Gaussian distribution with mean $\tau$ and variance $\sigma^{2}$, i.e., $\eta_{i}\sim \mathcal{N}\left(\tau,\sigma^{2} \right),\ \forall i\in \mathcal{U}$.
	
	Having these, our optimization problem in the ISC-HetNet can be formulated as
	\begin{equation}
	\begin{aligned}
	\mathbf{P1}:\ \max_{x,n} \quad & \sum_{i\in \mathcal{U}}\sum_{j\in \mathcal{B}}x_{ij}\eta_{i}S_{i}^{P}\left(b_{ij}\right)\\
	\textrm{s.t.} \quad & \sum_{j\in \mathcal{B}_{i}} x_{ij}= 1,\  \forall i\in \mathcal{U},\\
	&\sum_{i\in \mathcal{U}}x_{ij} n_{ij}\leqslant N_{j},\ \forall j\in \mathcal{B},\\
	& x_{ij}\in \left\{ 0,1\right\},\ \forall \left( i,j\right) \in \mathcal{U}\times \mathcal{B},
	\end{aligned}
	\end{equation}
	where $\mathcal{B}_{i}$ in the UA constraint denotes the set of feasible BSs storing the required KBs of $\mathcal{MU}_{i}$.
	The main difficulty in solving~$\mathbf{P1}$ lies on the stochasticity of $\eta_{i}$.
	It is precisely the intervention of random coefficients that turns~$\mathbf{P1}$ into a nondeterministic problem, which is also the biggest difference from that in traditional non-semantic communication system.
	
	\section{Solutions to UA and BA in the ISC-HetNet}
	In order to solve~$\mathbf{P1}$, we propose a two-stage method to reach the optimality.
	The first stage of our solution is to transform the original nondeterministic problem into a deterministic one, utilizing a stochastic programming model.
	Then we exploit an interior-point method and a heuristic algorithm in the second stage to find the solutions to UA and BA in the ISC-HetNet.
	
	\subsection{STM Maximization with Semantic Confidence Level}
	After carefully examining~$\mathbf{P1}$, we find that the random variables only exist in its objective.
	A possibility of handling such a problem is to take account of the distribution of the random objective function.
	According to Kataoka's model~\cite{kataoka1963stochastic}, we introduce a new objective function along with an extra constraint to make~$\mathbf{P1}$ suitable for stochastic programming without altering the original intention, by which the primal problem is now equivalent to
	\begin{equation}
	\begin{aligned}
	\mathbf{P1.1}:\ \max_{x,n} \quad & \bar{F}(x,n)\\
	\textrm{s.t.} \quad & \Pr\left\{F(\eta,x,n)\geqslant \bar{F}(x,n)\right\}\geqslant \alpha,\\
	&\sum_{j\in \mathcal{B}_{i}} x_{ij}= 1,\  \forall i\in \mathcal{U},\\
	&\sum_{i\in \mathcal{U}}x_{ij} n_{ij}\leqslant N_{j},\ \forall j\in \mathcal{B},\\
	& x_{ij}\in \left\{ 0,1\right\},\ \forall \left( i,j\right) \in \mathcal{U}\times \mathcal{B},\label{P2.1}
	\end{aligned}
	\end{equation}
	where we put
	\begin{equation}
		F(\eta,x,n)=\sum_{i\in \mathcal{U}}\sum_{j\in \mathcal{B}}x_{ij}\eta_{i}S_{i}^{P}\left(b_{ij}\right).
	\end{equation}
	The above transformation strictly keeps the same optimality of the original problem~$\mathbf{P1}$.
	Specifically, the aim of this problem becomes to maximize the lower bound $\bar{F}(x,n)$ defined by the newly introduced probabilistic constraint, in which $\Pr\{\cdot\}$ is the probability measure and $\alpha$ is a prescribed ($0<\alpha <1$, large in practice) confidence level.
	Generally, we can reckon $\alpha$ as a semantic confidence level or satisfaction of STM that can be achieved in the ISC-HetNet.
	
	Then, we notice that $F(\eta,x,n)$ has a nondegenerate distribution in the case of the optimal solution, such that
	\begin{equation}
		\Pr\left\{F(\eta,x,n)\geqslant \bar{F}(x,n)\right\}=\alpha,\label{Probaconstraint}
	\end{equation}
	in which the value of $\bar{F}(x,n)$ could not be increased to a constant without violating the constraints in~(\ref{P2.1}).
	By applying Theorem 10.4.1 given in~\cite{prekopa2013stochastic} to~$\mathbf{P1.1}$, we thus obtain
	\begin{equation}
		\begin{aligned}
			\bar{F}(x,n)=\ & \tau\sum_{i\in \mathcal{U}}\sum_{j\in \mathcal{B}}x_{ij}S_{i}^{P}\left(b_{ij}\right)\\
			& -\sigma\Phi^{-1}(\alpha)\sqrt{\sum_{i\in \mathcal{U}}(\sum_{j\in \mathcal{B}}x_{ij}S_{i}^{P}(b_{ij}))^{2}},
		\end{aligned}
	\end{equation}
	where $\Phi^{-1}(\cdot)$ is the inverse function of the standard normal probability distribution.
	Further, we relax the association constraint by $0\leqslant x_{ij}\leqslant1$, and fix $n_{ij}=n_{ij}^{T}$ to satisfy a minimum bit-rate requirement based on ISC quality.
	To this end, we have a deterministic maximization problem by rewriting~$\mathbf{P1.1}$ to
	\begin{equation}
		\begin{aligned}
		\mathbf{P1.2}:\ \max_{x} \quad & \bar{F}(x)\\
		\textrm{s.t.} \quad & \sum_{j\in \mathcal{B}_{i}} x_{ij}=1,\  \forall i\in \mathcal{U},\\
		&\sum_{i\in \mathcal{U}}x_{ij} n_{ij}^{T}\leqslant N_{j},\ \forall j\in \mathcal{B},\\
		& 0\leqslant x_{ij}\leqslant 1,\ \forall \left( i,j\right) \in \mathcal{U}\times \mathcal{B},
		\end{aligned}
	\end{equation}
	where
	\begin{equation}
		\bar{F}(x)=\tau\sum_{i\in \mathcal{U}}\sum_{j\in \mathcal{B}}x_{ij} \xi_{ij}^{T} -\sigma\Phi^{-1}(\alpha) \sqrt{\sum_{i\in \mathcal{U}}(\sum_{j\in \mathcal{B}}x_{ij}\xi_{ij}^{T})^{2}}.
	\end{equation}
	Here, $\xi_{ij}^{T}$ is a constant given by the fixed $n_{ij}^{T}$.
	Based on the convexity proof in~\cite{kataoka1963stochastic} and~\cite{prekopa2013stochastic}, it can be concluded that $\bar{F}(x)$ in~$\mathbf{P1.2}$ also holds convexity, and in practical, $\alpha>1/2$ is usually fulfilled so that we also have $\Phi^{-1}(\alpha)>0$.
	Note that $\tau$, $\sigma$, and $\alpha$ are all treated as constants as well, which values are fixed in the initial UA stage.
	
	\subsection{Optimal UA Solutions under Relaxation}
	For effective optimization, we utilize an interior-point method, i.e., the barrier method~\cite{potra2000interior}, as the solution to~$\mathbf{P1.2}$.	
	Technically, let $\varphi(x)$ be the logarithmic barrier associated with the bandwidth constraint, such that
	\begin{equation}
		\varphi(x)=\sum_{j\in \mathcal{B}}\log(N_{j}-\sum_{i\in \mathcal{U}}x_{ij} n_{ij}^{T}).
	\end{equation}
	Then the primal problem can be rephrased, making the inequality constraint implicit in the following objective
	\begin{equation}
	\begin{aligned}
	\mathbf{P1.3}:\ \max_{x} \quad & W(x,r)\\
	\textrm{s.t.} \quad & \sum_{j\in \mathcal{B}_{i}} x_{ij}=1,\  \forall i\in \mathcal{U},\\
	& 0\leqslant x_{ij}\leqslant 1,\ \forall \left( i,j\right) \in \mathcal{U}\times \mathcal{B},
	\end{aligned}
	\end{equation}
	where
	\begin{equation}
			W(x,r)=\bar{F}(x)+r\cdot \varphi(x).
	\end{equation}
	Here $r$ is a small positive scalar that sets the accuracy of the approximation, and as $r$ decreases to zero, the maximum of $W(x,r)$ should converge to a solution of~$\mathbf{P1.2}$.
	
	It is important to mention that $W(x,r)$ still holds convexity since both $\bar{F}(x)$ and $\varphi(x)$ are convex.
	Given $r>0$, we can directly find its central points $\textbf{X}^{*}(r)$, which are point sets containing the optimal $x_{ij}^{*}(r)$, $\forall \left( i,j\right) \in \mathcal{U}\times \mathcal{B}$.
	By iterating over the descent value of $r$, we eventually obtain the optimal $\textbf{X}^{*}$ of~$\mathbf{P1.2}$.
	However, each association variable in $\textbf{X}^{*}$ cannot guarantee a binary value (i.e., $0$ or $1$), therefore, a further approximation approach is needed for the final UA and BA.
	
	\subsection{Association Approximation and Allocation Adjustment}
	In this section, we heuristically require each MU to approximate its association variables in $\textbf{X}^{*}$ to either $1$ or $0$, which should follow the below rule as
	\begin{equation}
		\label{projection}
			x_{ij}=\left\{\begin{aligned}
			1,\quad &  \text{if}\ j=\arg \max_{j\in \mathcal{B}_{i}}x_{ij}^{*}\\
			0,\quad &  \text{otherwise}
		\end{aligned}
		,\  \forall i\in \mathcal{U}.\right.
	\end{equation}
	 
	 Nevertheless, the resulted bandwidth consumption may exceed some BSs' budget after~(\ref{projection}).
	 In this regard, we choose to reassign the MUs who are consuming the most bandwidth resources of such BSs to other BSs, based on those MUs' weights list given in $\textbf{X}^{*}$, until the resource constraints of all BSs are satisfied.
	 At this point, we should further adjust our allocation strategy of the remaining bandwidth, denoted by $n_{ij}^{R}$, after UA phase is completed.
	 With the obtained $x_{ij}^{*}$ and the given $n_{ij}^{T}$,~$\mathbf{P1.2}$ can be simplified to ($\forall j\in \mathcal{B}$)
	 \begin{equation}
		\begin{aligned}
		\mathbf{P1.4}:\ \max_{n} \quad & \tau\sum_{i\in \mathcal{U}_{j}} S_{i}^{P}(b_{ij})\\
			& -\sigma\Phi^{-1}(\alpha)\sqrt{\sum_{i\in \mathcal{U}_{j}}(S_{i}^{P}(b_{ij}))}\\
		\textrm{s.t.} \quad &\sum_{i\in \mathcal{U}_{j}}n_{ij}^{T}+n_{ij}^{R}=N_{j},
		\end{aligned}
	\end{equation}
	where the set of MUs serving by $\mathcal{BS}_{j}$ is
	\begin{equation}
		\mathcal{U}_{j}=\left\{ i\mid x_{ij}=1\right\}.
	\end{equation}
	As~$\mathbf{P1.4}$ is a convex optimization problem, we can easily find the optimal solution of $n_{ij}^{R}$.
	After this, both UA and BA have been well optimized in the ISC-HetNet, while realizing high STM with relatively low computational complexity.
	
	\section{Numerical Results}
	In this section, we evaluate the performance of the proposed two-stage solution in an ISC-HetNet.
	Let $5$ PBSs, $10$ FBSs, and $200$ MUs randomly locate in a circular area with a radius of $500$ m, where the MBS is placed at the center.
	Besides, the transmit power of MBSs, PBSs, and FBSs is set to $43$ dBm, $35$ dBm, and $20$ dBm, respectively, and each BS has a bandwidth budget of $2$ MHz.
	Here we use $L(d)=34+40log(d)$ and $L(d)=37+30log(d)$ as the path loss model for MBSs/PBSs and FBSs, respectively, under a noise power of $-111.45$ dBm~\cite{boostanimehr2014unified}.
	For the ISC model part, we put all MUs into a wireless text communication environment, in which the Transformer model is employed, to realize $S_{i}^{P}(\cdot)$ for each $\mathcal{MU}_{i}$.
	Meanwhile, we set a mean value of $\tau=0.5$ and a variance value of $\sigma=0.1$ for each knowledge matching coefficient $\eta_{i}$, while having a semantic confidence of $\alpha=95\%$ in the two-stage solution.
	Furthermore, two baseline algorithms are utilized for comparisons, including~\textit{max-SINR plus water-filling} and~\textit{max-SINR plus evenly-distributed}.
	Among them, each MU is only associated with the strongest BS in the UA phase, following by two classical BA algorithms, respectively, i.e., water-filling and even distribution.
	In addition, a bit-rate threshold of $0.01$ Mbit/s is fixed in our solution as well as the benchmarks to guarantee a basic service quality for users.
	
\begin{figure}[t]
		\centering
		\includegraphics[width=0.49\textwidth]{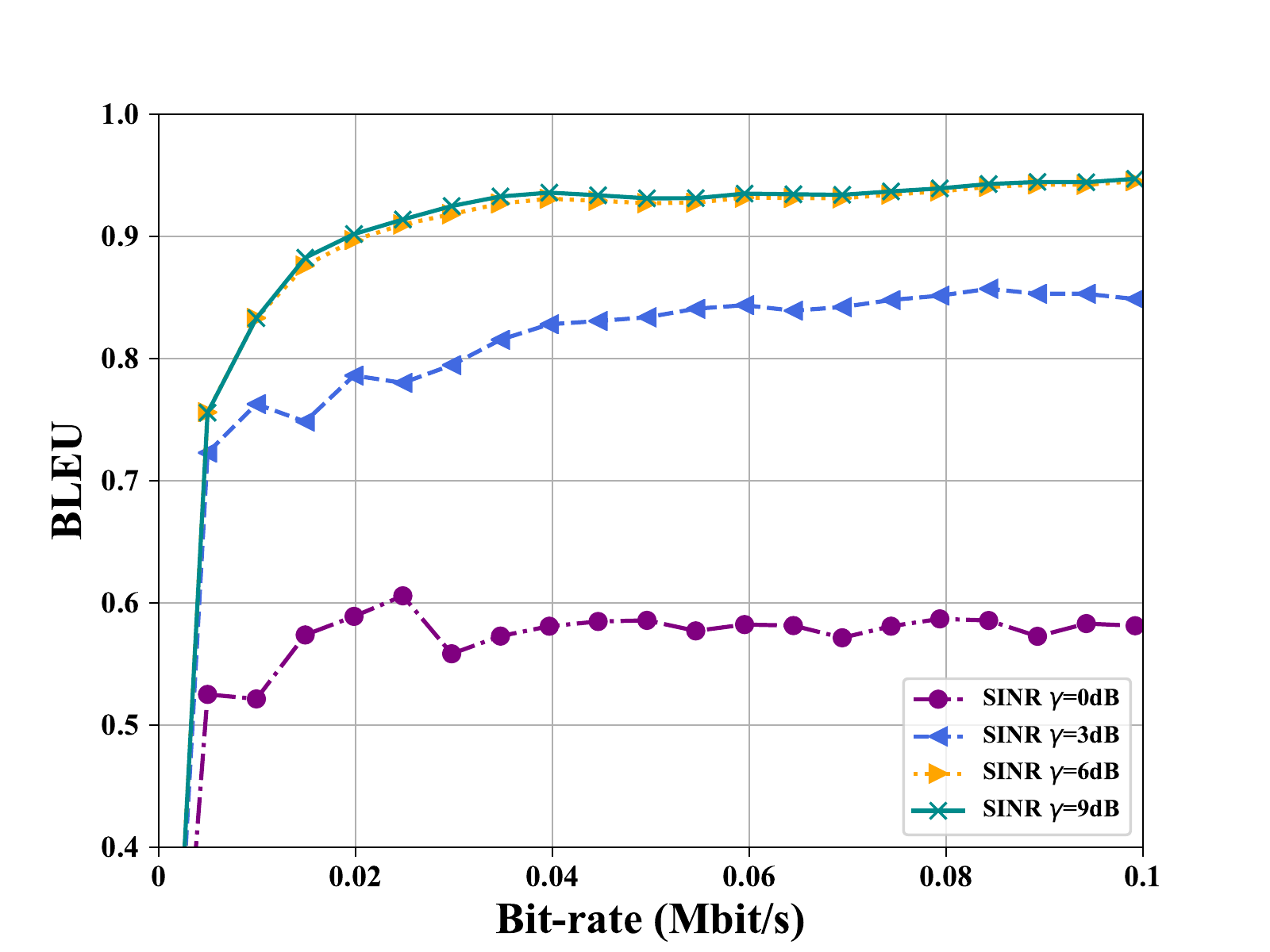} 
		\caption{The BLEU score (1-gram) vs. bit-rates obtained by MUs in the ISC-HetNet under four different SINRs of $0$, $3$, $6$, and $9$ dB.}
		\label{BLEU}
    \end{figure}

	Fig.~\ref{BLEU} shows the results of bilingual evaluation understudy (BLEU) scores ($1$-gram) versus different bit-rates under four different SINRs in the channel.
	Notably, BLEU is a classical metric in NLP by counting the difference of words between input and output texts, and the closer its score is to 1, the higher the text similarity~\cite{papineni2002bleu}.
	It can be seen that the BLEU under each SINR first grows as the bit-rate improves, and will soon stay at a stable score after about $0.03$ Mbit/s.
	Likewise, we observe a higher BLEU when the SINR increases from $0$ dB, and its score will stabilize at almost the same high level around $0.93$ after $6$ dB.
	Both phenomena indicate the necessity of providing a minimum bit-rate for MUs under good channel conditions, so as to achieve low-error semantic communication.
	In this context, if a stable high accuracy for semantic decoding can be ensured, it is believed that each MU can obtain a higher message-rate as the bit-rate improves, while its growth rate can also keep at a stable level related to its BLEU.
	This conclusion sufficiently validates the properties of the B2M function we defined in ISC, such as its convexity or monotonically increasing nature, etc.
	
    \begin{figure}[t]
		\centering
		\includegraphics[width=0.49\textwidth]{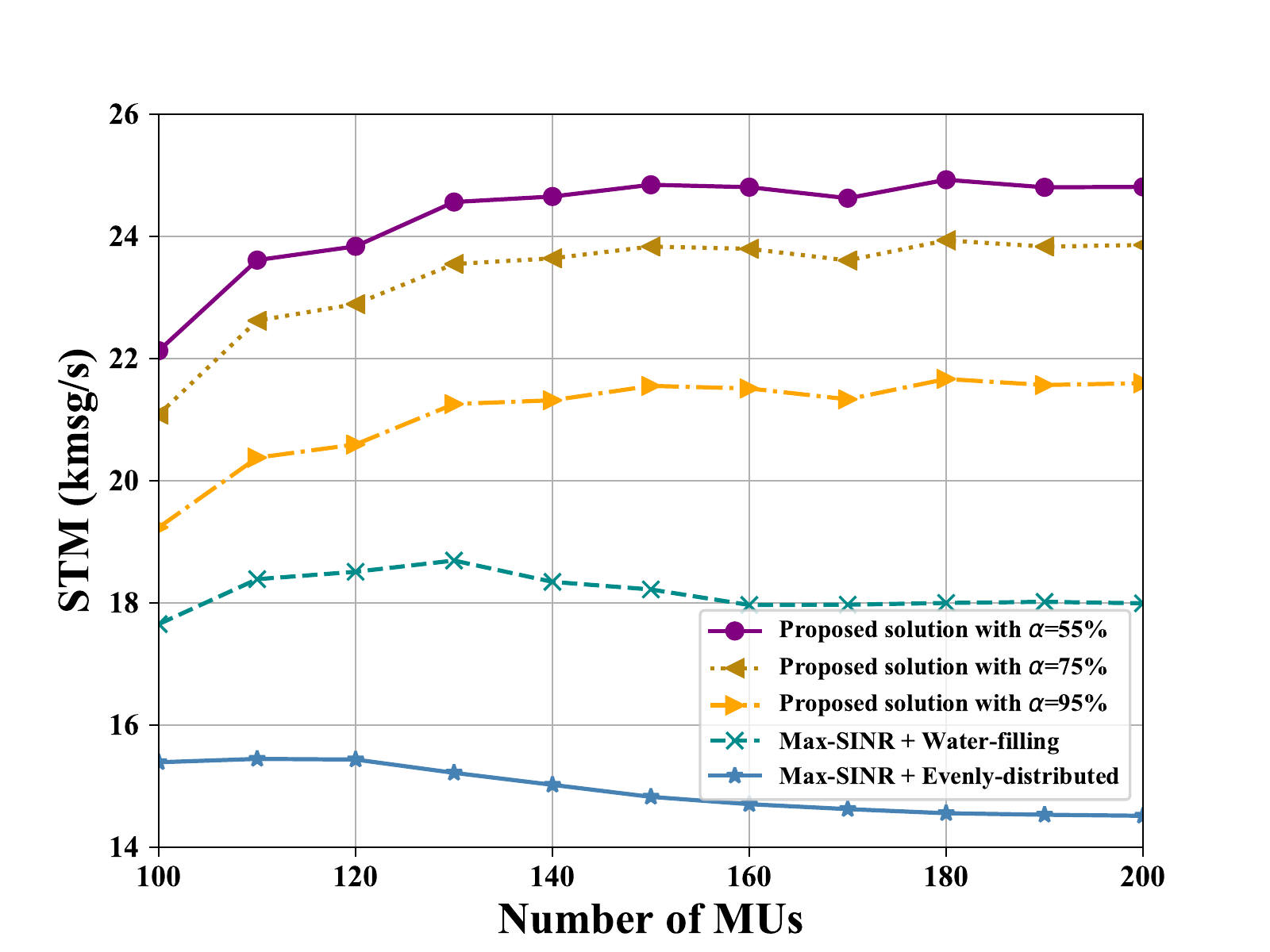} 
		\caption{The STM performance against number of MUs under three different semantic confidence levels of $\alpha=55\%$, $75\%$, and $95\%$.}
		\label{confi}
    \end{figure}
    
    \begin{figure}[t]
		\centering
		\includegraphics[width=0.49\textwidth]{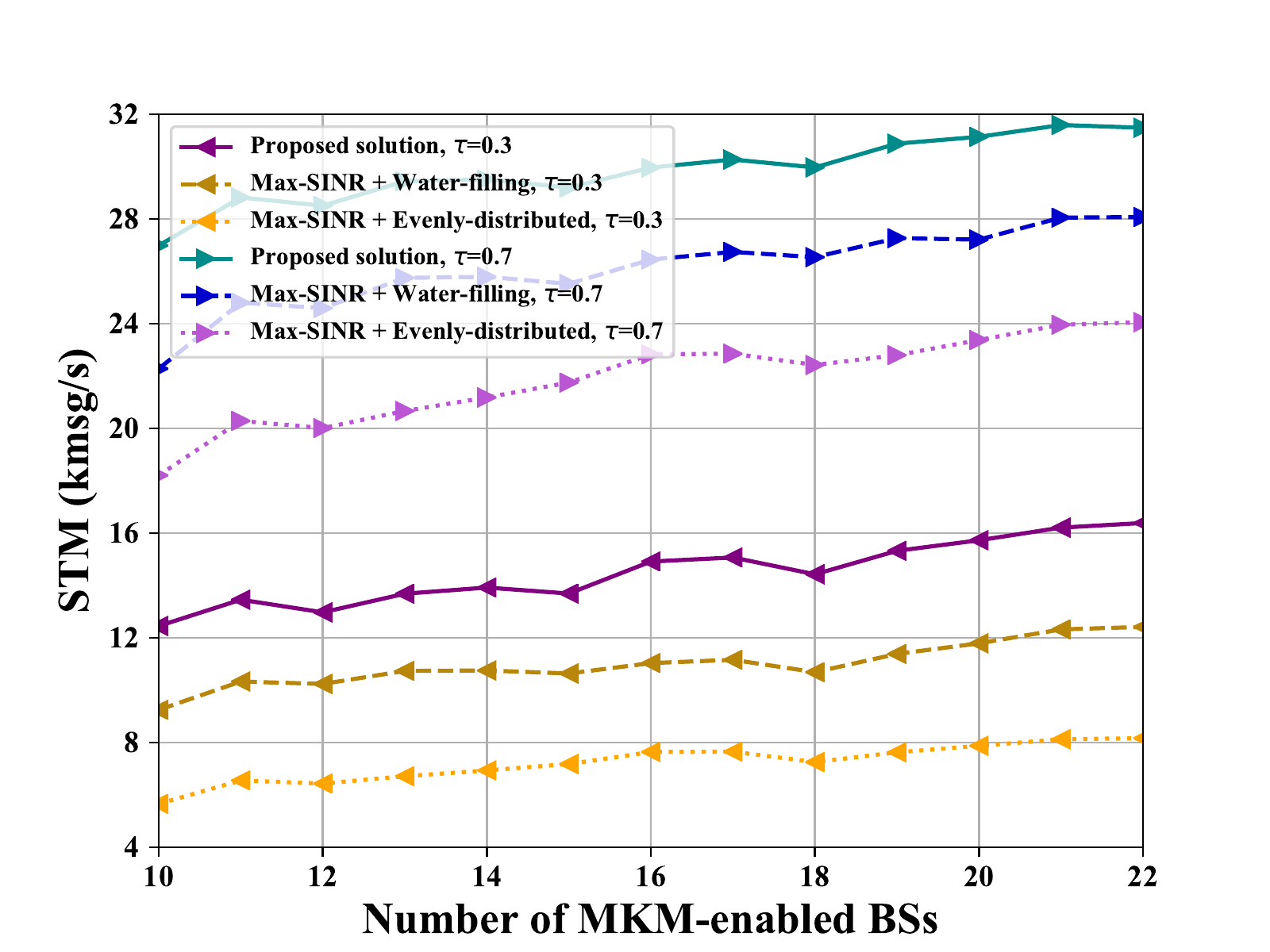} 
		\caption{The STM performance against number of KB-enabled BSs under two different mean values of $\tau=0.3$ and $0.7$ for knowledge matching coefficients.}
		\label{mean}
    \end{figure}
		
	Next, we compare the STM against different numbers of MUs in Fig.~\ref{confi}, where the results under three different $\alpha$ in our first-stage solution are also rendered.
	Specifically, the STM seen by our solution increases with the number of MUs at the beginning, and then soon remains at a stable performance after about $130$ MUs.
	This is because the bandwidth budgets of some BSs start to be reached after the number of MUs exceeds $130$, so that the whole STM is inevitably stabilized based on our UA adjustment strategy.
	Besides, an upward trend of STM performance is observed along with the reduction of the preset semantic confidence level.
	Since a higher semantic confidence represents a lower allowable limit as in~(\ref{Probaconstraint}), this can apparently lead to a worse STM result.
	Moreover, our solution consistently outperforms the two benchmarks, even at the highest confidence level of $\alpha=95\%$.
	
	At last, we present the STM obtained at different numbers of KB-enabled BSs under two mean values preset for the stochastic parameter $\eta_{i}$, as shown in Fig.~\ref{mean}.
	In this figure, we can see a slow growth trend of STM in all methods as the number of BSs increases, which is because there are more available resources in the ISC-HetNet and the likelihood of having overloaded BSs decreases.
	In parallel, as expected that a higher mean value of $\eta_{i}$ enables a higher STM.
	This result reveals a fact that the ISC-HetNet can achieve a better STM when there is a higher knowledge matching degree between each MU and its associated BS.
	Furthermore, our solution can also surpass the two max-SINR algorithms on STM under different mean values.
	To summarize, both UA and BA of the ISC-HetNet can be well addressed by the proposed two-stage solution, showing the great superiority and reliability when compared to the max-SINR-based algorithms.
	
	\section{Conclusions}
	This paper addressed both UA and BA problems in an ISC-HetNet with an STM-maximization objective.
	Under the premise of fully combining the characteristics of ISC, we built a semantic network model by introducing the auxiliary KB and a new semantic-based performance metric.
	Afterward, a corresponding optimization problem was formulated when considering the dynamic background knowledge matching condition.
	To cope with this problem, we specially proposed a two-stage solution.
	Compared with the two baseline algorithms, the numerical results presented that a higher STM performance can be reached by the proposed solution with sufficient reliability.
	Therefore, this paper is expected to be a promising work to provide inspirations on intelligent semantic communication from a networking perspective.
	
	\bibliographystyle{IEEEtran}
	\bibliography{main}
\end{document}